# Compositional Analysis of the High Molecular Weight Ethylene Oxide Propylene Oxide Copolymer by MALDI Mass Spectrometry

Orwa Jaber Houshia[1] & Charles Wilkins[2]

[1] Department of Chemistry, Arab American University, Jenin-WestBank, Palestine

[2] Department of Chemistry and Biochemistry, University of Arkansas, 345 N. Campus Drive, Fayetteville, AR 72701, USA

Correspondence: Orwa Jaber Houshia, Department of Chemistry, Arab American University, PO box 240, Jenin-WestBank, Palestine. Tel: 972-42-510-801. E-mail: orwa.housheya@aauj.edu

*The research was funded by the National Science Foundation grants CHE-00-91868, CHE-99-82045, and CHE-04-5513)*

## Abstract

The composition of narrow distribution poly ethylene oxide-propylene oxide copolymer (Mw ~ 8700 Da) was studied using matrix assisted laser desorption ionization (MALDI) mass spectrometry. The ethylene oxide-propylene oxide copolymer produced oligomers separated by 14 Da. The average resolving power over the entire spectrum was 28,000. Approximately 448 isotopically resolved peaks representing about 56 oligomers are identified. Although agreement between experimental and calculated isotopic distributions was strong, the compositional assignment was difficult. This is due to the large number of possible isobaric components. The purpose of this research is to resolve and study the composition of high mass copolymer such as ethylene oxide-propylene oxide.

## 1. Introduction

The amphiphilic pluronic copolymers, ethylene-oxide propylene-oxide (EO-PO), are of high commercial importance and are used as nonionic surfactants for numerous applications in the pharmaceutical, biomedical, and chemical industries (Kalinoski, 1996; Priorr, 1987; Ryan & Stanford, 1989). These biodegradable and biocompatible copolymers have the ability to form micelles and hydrogels in water because they consist of a hydrophilic segment, ethylene-oxide, and a hydrophobic segment, propylene-oxide (Alexandridis & Hatton, 1995; Chu, 1995; Gref et al., 1994; Hamley, 1998; Jeong, Bae, Lee, & Kim, 1997; Li, Rashkov, Espartero, Manolova, & Vert, 1996; Schick & Fowkes, 1996; Tanodekaew, Pannu, Heatley, Attwood, & Booth, 1997).

By controlling and adjusting the composition of the EO/PO segments in the copolymer, the physical properties of these surfactants can be engineered to fit a desirable application. Because many of the macroscopic physical properties depend on the composition, accurate determination of the composition is an essential part of the copolymer design. A number of techniques are available to characterize the copolymers including NMR, IR, GPC, Raman, and Viscosimetry. These tools give good structural and bulk compositional information about the copolymer. However, mass spectrometry can give detailed information (Maciejczek, Mass, Rode, & Pasch, 2010; Weidner & Falkenhagen, 2011).

Mass spectrometry has been extensively applied to copolymer analysis, especially since the matrix-assisted laser desorption/ionization (MALDI) technique was introduced. MALDI mass spectrometry allows for direct elucidation of end groups, molecular weight, and the fine compositional details simultaneously (Montaudo, 1999, 2001, 2002; Montaudo & Samperi, 1998; Przybilla, Francke, Rader, & Mullen, 2001; Servaty et al., 1998; van Rooij et al., 1998; Wilczek-Vera, 1996; Wilczek-Vera, Yu, Waddell, Danis, & Eisenberg, 1999a, 1999b; Yoshida, Yamamoto, & Takamatsu, 1998; Yu, Vladimirov, & Frechet, 1999; Zoller & Johnston, 2000). A single spectrum is often sufficient to give complete information about the copolymer (end groups, repeat units, molecular weight, and compositional details). In particular, MALDI Fourier transform mass spectrometry (FTMS) is able to distinguish closely spaced peaks-with isobar separation (Kaufman, Jaber, Stump, Simonsick, & Wilkins, 2004). MALDI mass spectrometry has become a popular tool for characterizing polymers and copolymers for several reasons. The MALDI process predominantly produces singly charged species, making spectral results simple and easy to interpret. The desorbed high molecular weight ions generally survive desorption without major fragmentation. Challenging nonpolar hydrocarbon polymers such as polyethylene (Jaber & Wilkins, 2005) can be analyzed which might be difficult or impossible with electrospray source ionization (ESI) techniques. Finally, one can take advantage of solventless MALDI preparation methods for insoluble polymers (Dolan & Wood, 2004; Marie, Fournier, & Tabet, 2000; Pruns et al., 2002; Przybilla, Brand, Yoshimura, Rader, & Mullen, 2000; Skelton, Dubois, & Zenobi, 2000; Trimpin, Grimsdale, Rader, & Mullen, 2002; Trimpin, Rouhanipour, Az, Räder, & Müllen, 2001).

The complimentary technique to MALDI mass spectrometry, ESI, can generate a wealth of information about polymers and copolymers (O'Connor & McLafferty, 1995; Shi, Hendrickson, Marshall, Simonsick, & J., 1998). But by far, the most frequent mass spectrometry tool used for polymer and copolymer characterization is MALDI time-of flight (TOF) mass spectrometry (Hanton, 2001; Nielen, 1999; Rader & Schrepp, 1998). However, the limited mass resolution of TOFMS makes it inadequate and unreliable for compositional analysis of copolymer of high molecular weight. Several EO-PO copolymer analyses have been reported for masses below 5000 Da (Chen, Zhang, Tseng, & Li, 2000; Jaber & Wilkins, 2005; Schriemer & Li, 1996; Terrier, Buchmann, Cheguillaume, Desmazieres, & Tortajada, 2005; van Rooij et al., 1998). Here, we show that MALDI-FTMS extends the compositional EO-PO copolymer analysis for oligomers with masses up to 9900 Da. FTMS provides the resolving power needed to isotopically resolve the oligomers. However, even with high resolution spectra, interpretation of the results is difficult at high masses due to the large number of possible isobaric components. Thus, a correlation function was implemented to match the experimental isotopic distribution with the theoretical one in order to narrow the possible compositional choices. It is worth mentioning molecular weights up to 1.5 million Dalton have been detected by MALDI-TOF, but neither oligomer resolution nor isotopic resolution was reported (Schriemer & Li, 1996).

## 2. Experimental Methods

### 2.1 Material and Sample Preparation

Text Poly (ethylene oxide-propylene oxide, mw ~8700) copolymer was purchased from Polysciences, Inc. (Warrington, PA). NaCl was purchased from Aldrich Chemical Co. (Milwaukee, WI). The matrix, 2,5-dihydroxybenzoic acid (DHB), was obtained from Fluka (Milwaukee, WI). Methanol was obtained from EM Science (Gibbstown, NJ). All reagents were used without further purification. DHB was dissolved in methanol to make a 0.5 M solution. NaCl was dissolved in water at a concentration of $1.15 \times 10^{-3}$ M. The copolymer was dissolved in methanol to a concentration of 5.0 mg/ml Deposition of sample on the probe tip was done as follows: 1.0 μL of copolymer sample is added and allowed to air dry, followed by addition of 0.2 μL of NaCl solution and a final layer of 2.0 μL of DHB matrix solution added. The sample is allowed to air dry.

### 2.2 Instrumentation

MALDI experiments were performed using both a 9.4 Tesla Fourier transform mass spectrometer and a reflectron time-of-flight mass spectrometer. The FTMS instrument is an IonSpec Ultima (Lake Forest, CA) with an external ionization source, utilizing a 9.4-Tesla superconducting magnet. The FTMS is also equipped with an ESI source. A Bruker Reflex III reflectron TOF (Billerica, MA) is used for the MALDI-TOF measurements. A 355 nm pulsed Nd-YAG laser (New Wave, Inc.) was used with FTMS. For The TOF measurements a 337 nm pulsed laser N2 laser was used. For the Ultima FTMS, ions are created externally by MALDI and guided into the ICR cell using RF quadrupole ion optics. FTMS spectra were acquired in positive ion mode at a pressure of ~ 5 x $10^{-10}$ Torr. Each spectrum resulted from pulsing the laser 7 times at the same spot on the sample. Ions were gated into the trapping cell after every 7th laser pulse. Spectra measured with the Bruker Reflex III TOF were measured using a 337 nm laser while operating in reflectron mode. Each TOF spectrum resulted from 200 laser pulses. The data analysis was handled by an in-house computer software-program copolymer calculator.

### 2.3 Data Processing

A home-made computer program was created to calculate the possible copolymer compositions based on Microsoft EXCEL platform and executed using a Pentium 4 1.6 GHz computer equipped with 768 MB of RAM memory. The program let us select multiple end groups, multiple cation attachment and up to three copolymer segments. The program calculates all of the possible compositional copolymer combinations and then searches the database of mass tables within a specified mass accuracy interval usually within ± 50 ppm.

## 3. Results and Discussion

There have been a limited number of reports on compositional analysis of high molecular weight (>5000 Da) copolymers by MALDI mass spectrometry. This deficiency derives from the fact that many MS instruments are limited in mass range. Furthermore analyses are often performed by relatively low resolution TOFMS. For example, Figure 1 is a MALDI-TOF spectrum of EO-PO copolymer (mw ~8700) acquired in reflectron positive ion mode. The spectrum was the result of the sum of 200 laser pulses on the MALDI target. The inset within Figure 1 shows an expanded mass range from m/z 8820-8880. The average resolving power was ~400. Due to this low resolving power, oligimeric resolution was difficult to observe and isotopic resolution impossible. Upon attempting to analyze this copolymer using ESI-FTMS, no signal was detected. The reasons for this are not clear. So, in the present study, the alternative of MALDI-FTMS was employed. Mass peak assignment is easier when spectra are acquired at high resolving power. High field FTMS is capable of yielding isotopically resolved peaks of polymers with masses up to m/z 12,000 Da (Jaber & Wilkins, 2005). The MALDI-FTMS spectrum shown in Figure 2 was acquired with a 9.4 Tesla FTMS. The spectrum was the result of 7 laser shots on the same spot to generate ions, which were gated into the cell after the 7th pulse. A single transient was transformed after excitation of ions to result in a single spectrum. Under these conditions, the average resolving power was about 28000. The mass envelope of the spectrum extends from m/z 8300 to 9900. The mass values of the oligomers detected correspond to $(M+Na)^+$, where M is HO-$[(C2\ H_4O)m-(C_3H_6O)n]$-H. The fully resolved isotopic peaks in Figure 3B shows oligomers are separated by 14 mass units. The separation between the oligomers is due to the replacement of one EO unit by one PO unit: $(PO)_{y+1} + (EO)_{x-1}$. Mass assignment was difficult even though the peaks were well resolved. To speed up the data analysis, a computer software program copolymer calculator was used. Multiple possible compositional combinations were found for each peak. This was not unexpected since the number of isobaric structures increase as mass increases. In the program, possible compositional combinations that had mass accuracy within ± 40 ppm were accepted and those outside this interval were rejected. To narrow down the search, a correlation coefficient function (Equation 1) between the theoretical isotopic distribution (Figure 3A) and experimental results was implemented. Results with correlation coefficients between 0.8 and 1.0 were admissible provided that they had mass accuracy with ± 40 ppm.

$$r = \frac{SS_{xy}}{\sqrt{(SS_{xx})(SS_{yy})}} \tag{1}$$

Where

$$SS_{xy} = \frac{[\sum xy - (\sum x)(\sum y)]}{n}$$

$$SS_{xx} = \sum x^2 - \frac{(\sum x)^2}{n}$$

$$SS_{xx} = \sum y^2 - \frac{(\sum y)^2}{n}$$

Where x is the intensity from the experimental isotopic distribution, y is the intensity of the theoretical isotopic distribution and r is the correlation coefficient. Table 1 illustrates some of the multiple acceptable compositional combinations and their respective mass accuracies and correlation coefficients. Note that high correlation does not imply causality. The only valid conclusion is that a linear trend may exist between the theoretical and the experimental results, but it would incorrect to conclude that a change in "x", for example, causes a change in "y". Therefore what is being established here is the strength of agreement between theory and experiment.

Now the search has been narrowed down, is it possible to explicitly point to a definite compositional assignment? More detailed scrutiny yields a more revealing and complex picture. To illustrate this, consider the mass 8839.9069 Da in Table 1. This is the measured mass for the expanded oligomer

Table 1. The Multiple-isobaric possible compositional combinations and their respective mass accuracy and correlation coefficients

| Measured Mass | Theoretical Mass | PO units | EO units | Error (ppm) | r (correlation coefficient) |
|---|---|---|---|---|---|
| 8825.3771(A+1) | 8825.9571 | 99 | 69 | 9 | 0.93 |
| 8825.3771(A+1) | 8825.7963 | 77 | 98 | -9 | 0.93 |
| 8825.3771(A+1) | 8826.1179 | 121 | 40 | 27 | 0.92 |
| 8825.3771(A+1) | 8825.6355 | 55 | 127 | -27 | 0.93 |
| 8839.9069(A+1) | 8839.9728 | 100 | 68 | 8 | 0.96 |
| 8839.9069(A+1) | 8839.812 | 78 | 97 | -11 | 0.97 |
| 8839.9069(A+1) | 8840.1336 | 122 | 39 | 26 | 0.95 |
| 8839.9069(A+1) | 8839.6512 | 56 | 126 | -29 | 0.97 |
| 8854.9528(A+2) | 8854.9918 | 101 | 67 | 4 | 0.88 |
| 8854.9528(A+2) | 8854.831 | 79 | 96 | -14 | 0.88 |
| 8854.9528(A+2) | 8855.1526 | 123 | 38 | 23 | 0.87 |
| 8854.9528(A+2) | 8854.6702 | 57 | 125 | -32 | 0.88 |
| 8868.9548(A+2) | 8869.0074 | 102 | 66 | 6 | 0.82 |
| 8868.9548(A+2) | 8868.8466 | 80 | 95 | -12 | 0.83 |
| 8868.9548(A+2) | 8869.1682 | 124 | 37 | 24 | 0.81 |
| 8868.9548(A+2) | 8868.6858 | 58 | 124 | -30 | 0.83 |

Spectrum shown in Figure 4. For this mass the possible compositions are: PO100-EO68, PO78-EO97, PO122-EO39, PO56-EO126. For any of these isobaric compositions, a total of 17 theoretical isotopic peaks should be observed at resolving power of 28,000 as shown in Table 2. In this table the peak labeled "A" indicates that all carbons are the 12C (the monoisotopic peak). The "A+1" peak indicates that there is one 13C and the rest of the carbons in that peak are 12C, and so on, until "A+16". However, instead of observing 17 isotopic peaks only 12 peaks are seen as shown in the theoretically generated spectrum of Figure 5, due to the fact that some peaks are of very low intensity (see Table 2). Despite the striking similarities between the experimental (Figure 4) and the theoretical (Figure 5) isotopic distributions, intriguing differences are observed. First, the peaks in Figure 4 labeled with asterisks display higher intensity than their counterparts in Figure 5. Second, the shoulder peaks appear (overlapping-not well resolved) and are labeled with downward arrows in Figure 4. These observations raise an important question as to the source or the cause of these discrepancies between the two distributions. We postulate that the distorted experimental distribution in Figure 4 is a direct consequence of the coexistence of all 4 possible compositional outcomes. It is very likely that the presence of several possible compositional outcomes collectively contributing to this oligomer pattern. Confronted with uncertainty in choice of "best" combination, it is not surprising that "best" choice is a matter of probability. In general, the statistical probability of a particular result such as, $\alpha m \beta n$ for example, can be expressed in Equation 2 by the most probable outcome:

$$\rho(\alpha_m \beta_n) = \frac{(m+n)}{m!n!} \left( \frac{m}{m+n} \right)^m \left( \frac{n}{m+n} \right)^n \qquad (2)$$

Where m is the number of PO segments and n is the number of EO segments in the oligomer. As an example, starting with equal quantities of the monomers it can be shown that the most probable result for the combination PO100-EO68, PO78-EO97, PO122-EO39, PO56-EO126 is the one with PO56-EO126 compositional combination (this is the series labeled 182 in Figure 6). This means that

This choice is present at highest concentration and hence, it should seem reasonable that the other 3 possible isobaric compositions are at low concentration and have decreased detectability. Note that these isobaric compositions [POx-22 + EOy+29] differ by 0.16 mass units (Table 2) and it would take an average resolving power of 55000 to separate them. Their presence should not be ignored at all, as this would cause Figure 4 to exhibit dissimilarity to Figure 5. This may well be on the basis that one monomer is more reactive than the other.

## 4. Conclusion

Two new results are achieved here. First, MALDI copolymer mass analysis is shown to be feasible up to mass 9900 Da with meaningful results. Second, accurate mass measurement as a result of the high resolution is evident. However, compositional interpretation of high molecular weight copolymer is still complex, but MALDI-FTMS allows for direct measurements of oligomeric components. Such measurements of oligomeric species were impossible by MALDI-TOF with the resolution that was obtained. One obvious and critical issue for the MALDI-FTMS interpretation is the probable existence of multiple isobaric structures that contribute to the complexity in mass assignment. This issue is simplified with a combination of correlation functions and probability functions that help narrow down the search for the composition while retaining accurate information. Nevertheless, for extensive application to copolymers with high mass, the difficulties of analyzing compositions of higher mass copolymers will have to be resolved.

Table 2. An example of the theoretically generated isotopic intensity distribution for the mass starting at m/z 8839 Da

| (C3 H6 O)100 (C2 H4 O)68 | Intensity | (C3 H6 O)78 (C2 H4 O)97 | Intensity |
|---|---|---|---|
| A | 8838.9681 | 3.38 | A | 8838.8073 | 3.62 |
| A+1 | 8839.9719 | 16.5 | A+1 | 8839.8112 | 17.38 |
| A+2 | 8840.9757 | 41.38 | A+2 | 8840.8149 | 42.92 |
| A+3 | 8841.9793 | 70.86 | A+3 | 8841.8185 | 72.49 |
| A+4 | 8842.9827 | 93.14 | A+4 | 8842.8219 | 94.14 |
| A+5 | 8843.9861 | 100 | A+5 | 8843.8253 | 100 |
| A+6 | 8044.9890 | 91.32 | A+6 | 8844.8281 | 90.47 |
| A+7 | 8845.9923 | 72.87 | A+7 | 8845.8315 | 71.59 |
| A+8 | 8846.9955 | 51.77 | A+8 | 8846.8347 | 50.49 |
| A+9 | 8847.9983 | 33.26 | A+9 | 8847.8374 | 32.23 |
| A+10 | 8849.0019 | 19.55 | A+10 | 8848.841 | 18.84 |
| A+11 | 8850.0048 | 10.61 | A+11 | 8849.8437 | 10.18 |
| A+12 | 8851.0078 | 5.36 | A+12 | 8850.8468 | 5.12 |
| A+13 | 8852.0109 | 2.53 | A+13 | 8851.8499 | 2.41 |
| A+14 | 8853.0135 | 1.13 | A+14 | 8852.8525 | 1.07 |
| A+15 | 8854.0166 | 0.47 | A+15 | 8853.8556 | 0.45 |
| A+16 | 8855.0196 | 0.19 | A+16 | 8854.0595 | 0.18 |

| (C3 H6 O)122 (C2 H4 O)39 | Intensity | (C3 H6 O)56 (C2 H4 O)126 | Intensity |
|---|---|---|---|
| A | 8839.1288 | 3.15 | A | 8838.6465 | 3.88 |
| A+1 | 8840.1327 | 15.67 | A+1 | 8839.6504 | 18.31 |
| A+2 | 8841.1365 | 39.91 | A+2 | 8840.6541 | 44.51 |
| A+3 | 8842.1401 | 69.27 | A+3 | 8841.6577 | 74.16 |
| A+4 | 8843.1435 | 92.15 | A+4 | 8842.6611 | 95.14 |
| A+5 | 8844.147 | 100 | A+5 | 8843.6645 | 100 |
| A+6 | 8845.1498 | 92.19 | A+6 | 8844.6673 | 89.62 |
| A+7 | 8846.1532 | 74.18 | A+7 | 8845.6706 | 70.33 |
| A+8 | 8847.1564 | 53.1 | A+8 | 8846.6738 | 49.25 |
| A+9 | 8848.1592 | 34.33 | A+9 | 8847.6765 | 31.23 |
| A+10 | 8849.1628 | 20.3 | A+10 | 8848.6801 | 18.16 |
| A+11 | 8850.1655 | 11.07 | A+11 | 8049.6827 | 9.76 |
| A+12 | 8851.1688 | 5.62 | A+12 | 8850.6859 | 4.89 |
| A+13 | 8852.1719 | 2.66 | A+13 | 8851.6889 | 2.29 |
| A+14 | 8853.1745 | 1.19 | A+14 | 8852.6915 | 1.02 |
| A+15 | 8854.1777 | 0.5 | A+15 | 8853.6946 | 0.43 |
| A+16 | 8855.1806 | 0.2 | A+16 | 8854.6975 | 0.17 |

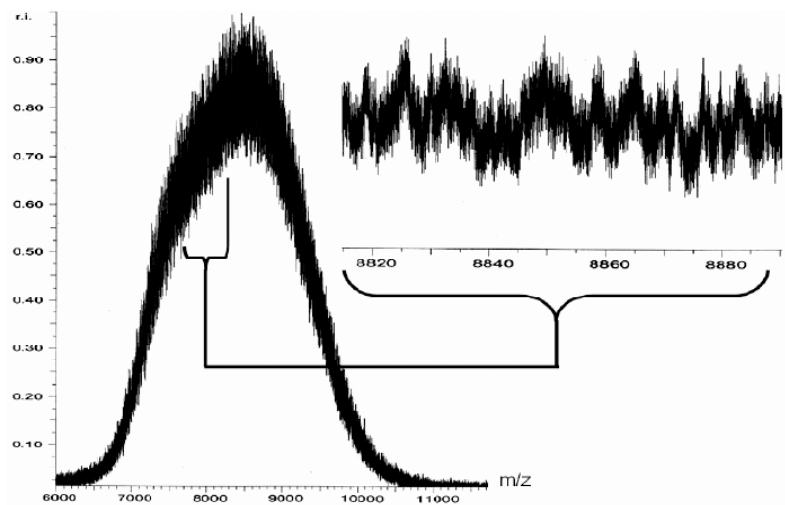

Figure 1. MALDI-TOF spectrum of the EO-PO copolymer with the inset showing an expanded mass range from m/z 8800 to 8890; Compared to that of the expanded FTMS spectrum

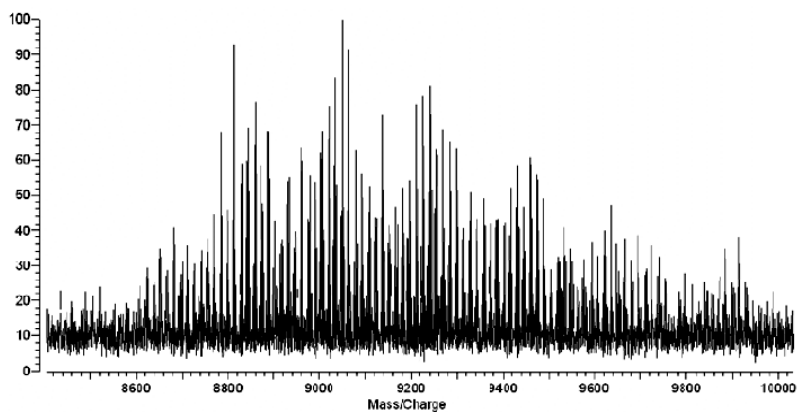

Figure 2. MALDI-FTMS of the EO-PO copolymer having a mass envelope from m/z 8300 to m/z 9900

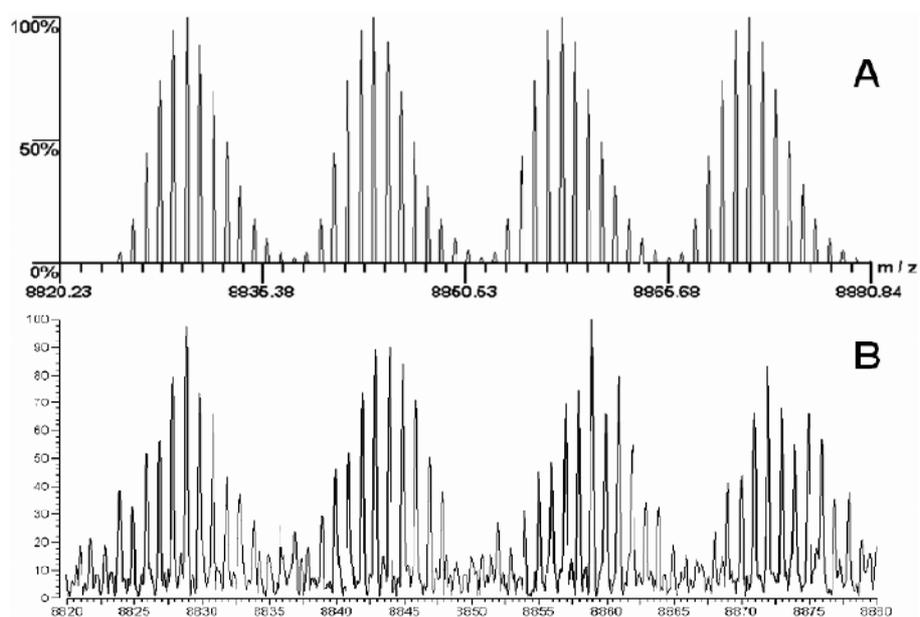

Figure 3. An expanded mass range of the MALDI-FTMS spectrum showing oligomeric and isotopic resolution: The top (A) spectrum is the theoretical range and the bottom (B) spectrum is the experimental result

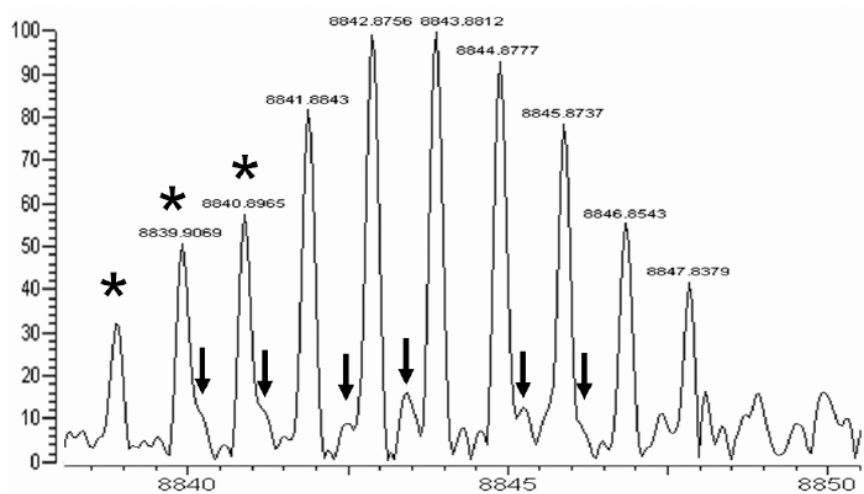

Figure 4. An isotopically resolved oligomer of the MALDI-FTMS spectrum. See text for explanation of the peaks with asterisks and arrows

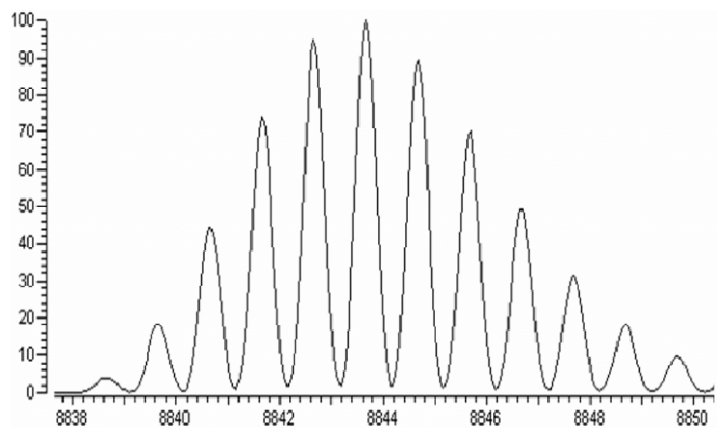

Figure 5. A theoretical isotopically resolved oligomer which corresponds to the experimental one shown in Figure 4

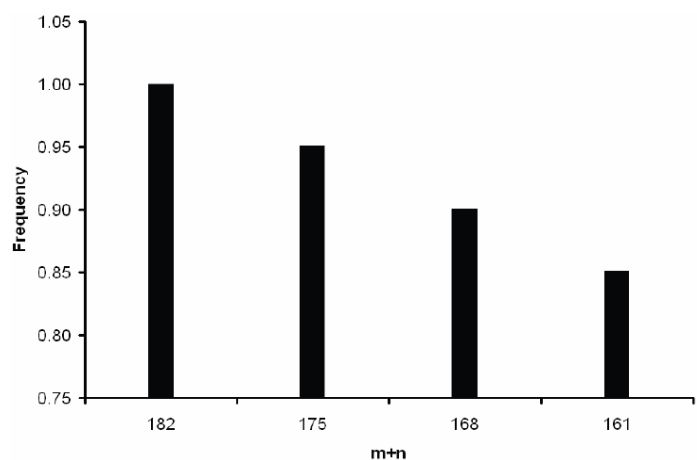

Figure 6. The most probable series of compositional combination of EO-PO segments

## Acknowledgments

The authors gratefully acknowledge support from National Science Foundation grants CHE-00-91868, CHE-99-82045, and CHE-04-55134.

**Notes**

Note 1. Orwa Jaber Housheya (aka, Arwah Jaber).